\documentclass[%
 reprint,
 amsmath,amssymb,
 aps,
]{revtex4-2}
\usepackage{bm, color}
\usepackage[colorlinks = true, allcolors = blue]{hyperref}
\usepackage{placeins}
\usepackage{float}

\usepackage{graphicx}
\usepackage{dcolumn}
\usepackage{bm}

\newcommand{\bx}{\mathbf{x}}

\begin{document}

\preprint{APS/123-QED}
\title{Dynamic Mode Decomposition for Extrapolating Non-equilibrium Green's Functions Dynamics}

\author{Cian C. Reeves}
\affiliation{%
Department of Physics, University of California, Santa Barbara, Santa Barbara, CA 93117
}%
\author{Jia Yin}%
\affiliation{%
Applied Mathematics and Computational Research Division, Lawrence Berkeley National Laboratory, Berkeley, CA 94720, USA}
\author{Yuanran Zhu}%
\affiliation{Applied Mathematics and Computational Research Division, Lawrence Berkeley National Laboratory,
Berkeley, CA 94720, USA}
\author{Khaled Z. Ibrahim}%
\affiliation{Applied Mathematics and Computational Research Division, Lawrence Berkeley National Laboratory,
Berkeley, CA 94720, USA}
\author{Chao Yang}
\affiliation{Applied Mathematics and Computational Research Division, Lawrence Berkeley National Laboratory,
Berkeley, CA 94720, USA}
\author{Vojt\ifmmode \check{e}\else \v{e}ch Vl\ifmmode \check{c}\else \v{c}ek}
\affiliation{%
Department of Chemistry and Biochemistry, University of California, Santa Barbara, Santa Barbara, CA 93117
}%
\affiliation{%
Department of Materials, University of California, Santa Barbara, Santa Barbara, CA 93117
}

\date{\today}
\begin{abstract}
The HF-GKBA offers an approximate numerical procedure for propagating the two-time non-equilibrium Green's function(NEGF). Here, using the $GW$ self-energy, we compare the HF-GKBA to exact results for a variety of systems with long and short-range interactions, different two-body interaction strengths and various non-equilibrium preparations.  We find excellent agreement between the HF-GKBA and exact time evolution in models when more realistic long-range exponentially decaying interactions are considered.  This agreement persists for long times and for intermediate to strong interaction strengths. In large systems, HF-GKBA becomes prohibitively expensive for long-time evolutions.  For this reason, look at the use of dynamical mode decomposition(DMD) to reconstruct long-time NEGF trajectories from a sample of the initial trajectory.  Using no more than 16\% of the total time evolution we reconstruct the total trajectory with high fidelity.  Our results show the potential for DMD to be used in conjunction with HF-GKBA to calculate long time trajectories in large-scale systems.
\end{abstract}

\maketitle

\section{\label{Intro}Introduction}
Despite the relevance of non-equilibrium physics in many condensed matter systems\cite{RevModPhys.81.163,Le_Hur_2016} a robust and practical theoretical framework for studying these non-equilibrium systems is still lagging.  One popular approach to studying non-equilibrium systems is to use two-time non-equilibrium Green's functions(NEGFs)\cite{stefanucci2013nonequilibrium, Kadanoff}.  For a given system and set of initial conditions, the time evolution of the one particle NEGF can be computed using the Kadanoff-Baym equations(KBEs)\cite{Stan_2009}.  This time evolution is exact, given the exact self-energy is known, but in practice, the self-energy is an approximate quantity.

Unfortunately, the KBEs are known to suffer from two major issues. Firstly, the KBEs have been reported to reach artificial steady states that are not present in the exact time evolution\cite{von_Friesen_2009,Friesen_2010,PhysRevB.90.125111}. The KBEs are in principle an exact set of equations when the exact self-energy is used, therefore these artificial steady states must arise due to the self-energy approximations used.  In fact, when the full self energy is included in finite systems, such as Hubbard clusters, the summation of Feynmann diagrams contributing to the self-energy leads to many exact cancellations\cite{von_Friesen_2009}.  However, when only certain classes of diagrams are summed to infinite order, some of these cancellations no longer occur.  Unphysical terms in the self-energy can build up from self consistency on only certain subsets of diagrams.  The unphysical terms may resemble an artificial bath that leads to the formation of spurious steady states, as was pointed out and demonstrated in \cite{von_Friesen_2009}.The second major problem with the KBE approach is in the computational cost, specifically in the number of time propagation steps ($N_t$). When solving the KBEs, the two-time Green's function needs to be propagated at all points on a two-time grid, which leads to asymptotic computational scaling $O(N_t^3)$\cite{bonitz2015quantum}.  This makes the use of the KBEs impractical outside of small systems and beyond short propagation times.  

Because of this second issue, an approximate partial solution to the KBE, known as the Hartree-Fock generalized Kadanoff-Baym ansatz(HF-GKBA), is more commonly used for simulations of realistic problems\cite{PhysRevB.34.6933,Hermanns_2012,Hermanns_2012,PhysRevLett.128.016801}. With the HF-GKBA, only the KBE for the two-time Green's function at equal times is explicitly propagated.  The time off-diagonal elements are then reconstructed from the information on the time-diagonal.  The HF-GKBA has become widely used due to the speedup it offers over the full KBE, especially for long time propagation's.  The HF-GKBA has even been argued to improve over the full KBE by removing spurious steady states and artificial damping\cite{von_Friesen_2009} and has even been claimed to outperform KBE in reconstructing particle densities in a simple Hubbard chain\cite{PhysRevB.90.125111}.  However, the HF-GKBA only makes further approximations upon those already made in the KBE, and so any improvements are, at least in part, fortuitous.  

Although faster than full KBE, in its original formulation the HF-GKBA still retains the $O(N_t^3)$ scaling (except when used with the Second-Born self-energy). Recently, a reformulation of the HF-GKBA, known as the G1-G2 scheme, has offered a method for propagating NEGFs within the HF-GKBA with scaling $O(N_t)$\cite{PhysRevB.101.245101}.   While promising, the scheme still suffers from several drawbacks.  Firstly, the method expresses the single particle self-energy in terms of the two-particle Green's function, thus upfolding the problem onto a larger space. This upfolding leads to $O(N_s^6)$ numerical scaling in the system size, making it difficult to use the G1-G2 scheme in large systems or systems with multiple bands. A second issue the G1-G2 scheme faces is that the time propagation is known to be unstable for long times and/or strong couplings since certain consistency relations for the two-particle Green's function break down\cite{PhysRevB.105.165155}. In the G1-G2 scheme, removing this instability requires the repeated diagonalization of the two-particle Green's function, making it impractical for large systems and long time evolutions\cite{PhysRevB.105.165155}. The former issue is well suited to the family of stochastic methods \cite{PhysRevLett.113.076402,doi:10.1021/acs.jctc.7b00770,PhysRevB.106.165129} which provides a seemingly straightforward strategy to reduce the computational cost in the system size. This, however, comes at a price: the time evolution trajectory is fundamentally not stable over extended propagation time.

All of this means in large scale or strongly interacting systems even HF-GKBA alone may be a computationally intractable tool to study NEGF over long-time trajectories.

In this paper, we test the ability of using dynamical mode decomposition(DMD) to reconstruct NEGF trajectories from partial samples of the full trajectory.  For the case of large systems, this would be extremely advantageous since DMD is computationally much cheaper than HF-GKBA.  Additionally, in the cases where we can successfully fit DMD on data before HF-GKBA becomes unstable, the need to diagonalize the two-particle Green's function at every step is removed.  In both scenarios, DMD is a promising method of significantly speeding up the propagation of NEGF under HF-GKBA, especially for long propagation times in large scale realistic systems. For smaller systems where exact diagonalization is possible we compare these results to the HF-GKBA result.  Further, we numerically demonstrate that (seemingly counterintuitively) the approximate methodology agrees better for the more realistic long-range Hamiltonian forms. This is rationalized as the approximate self-energy applied is well suited for realistic weakly and moderately correlated systems in which long-range interactions present significant contributions.  This further motivates the use of DMD for creating computationally cheap long-time trajectories that match well with exact time propagation.

\section{Theory}\label{Theory}
\subsection{Model Systems}\label{model_systems}
\begin{figure}[h!]
    \centering
    \includegraphics[width=.4\textwidth]{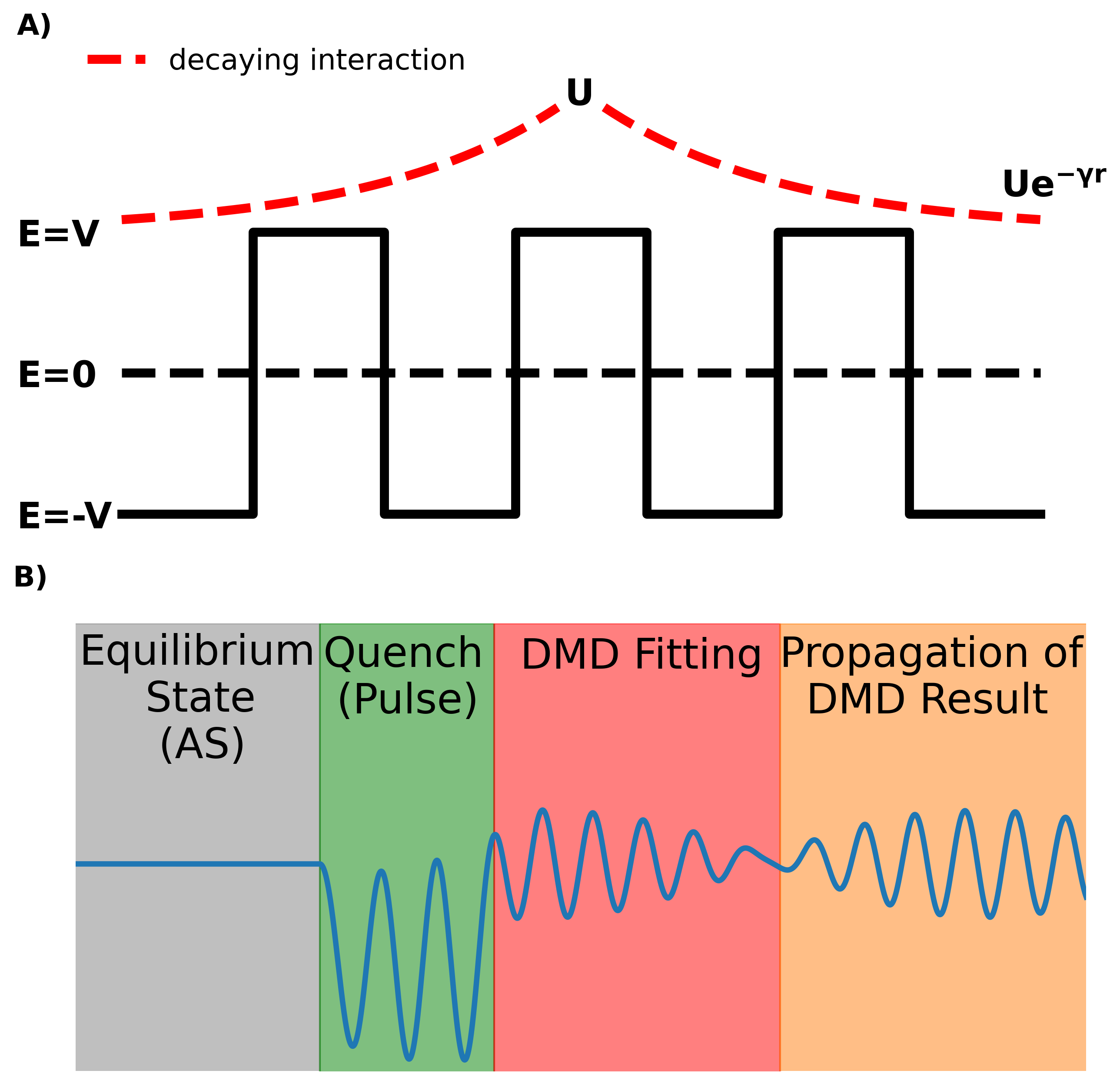}
    \caption{A) Model system with exponentially decaying interactions and an alternating local potential described by equations \eqref{MB_ham}, \eqref{two-body} and \eqref{single_particle}.  B) Outline of non-equilibrium preparation and DMD fitting procedure}
    \label{fig:sys_setup}
\end{figure}

To evaluate the ability of DMD to numerically extrapolate the Green's function trajectories, we use the following strategy illustrated in Fig \ref{fig:sys_setup}. The system is first prepared in a correlated stationary state and then driven from equilibrium via a quenching of specific sites of the Hamiltonian.  More details on the initial state preparation are given in \ref{sec:Methods}.  The discussion of system quenching is continued later in this section.  After the non-equilibrium dynamics are initiated, DMD is used in an initial window of the trajectory, after which the DMD result is propagated and compared with the remainder of the unfitted trajectory.  For a detailed discussion on the DMD procedure see section \ref{sub_sect:DMD}.
\FloatBarrier
\FloatBarrier
A generic many-body Hamiltonian can be written in the following form
\begin{equation}\label{MB_ham}
    \mathcal{H}  = \sum_{ij}h^{(0)}_{ij}(t)c^\dagger_ic_j + \frac{1}{2}\sum_{ijkl} w_{ijkl}(t) c^\dagger_ic_j^\dagger c_k c_l.
\end{equation}
Here $w_{ijkl}$ is the two-body interaction term and $h^{(0)}(t)$ is the single-particle Hamiltonian.  For the model we study, the corresponding two-body interactions is given explicitly by

\begin{equation}\label{two-body}
    \begin{split}
        w_{ijkl}^{\sigma_i\sigma_j\sigma_k\sigma_l}(t) &= U(t)\delta_{ij}\delta_{ik}\delta_{il}\delta_{\sigma_i\sigma_k}\delta_{\sigma_j\sigma_l}(1-\delta_{\sigma_i\sigma_k}) \\&\hspace{5mm}+ U(t)\sum_{n=1}^{N_s}\mathrm{e}^{-\gamma|i-j|}\delta_{ij}^{(n)}\delta_{ik}\delta_{jl}\delta_{\sigma_i\sigma_k}\delta_{\sigma_j\sigma_l},
    \end{split}
\end{equation}
where the $\sigma \in \{\uparrow,\downarrow\}$ are spin indices, $N_s$ is the number of sites in the chain and $\gamma$ determines the rate of decay of the long-range interactions. Here we define $\delta^{(n)}_{ij}$ to only be non-zero if $|i-j|=n$.

Later when using the HF-GKBA, since we prepare correlated initial states with adiabatic switching, we include an explicit time-dependence in the interaction terms above.  The initial state preparation will be discussed in more detail in section \ref{sec:Methods}, but for now we note that for a model with the two-body term in equation \eqref{two-body} and nearest neighbour hopping only the gap between the grounds state and the first excited state tends to $0$ as the system size is increased.  This makes the adiabatic switching procedure increasingly numerically unstable. To open a gap and allow for a numerically stable initial state preparation we add an alternating step potential to our model which is illustrated in Fig. \ref{fig:sys_setup}.  

The single particle Hamiltonian for our model is given by
\begin{equation}\label{single_particle}
    \begin{split}
        h^{(0)}_{ij}(t) &= -J\delta^{(1)}_{ij} +\delta_{ij}(-1)^i V + h^{\textrm{quench}}_{ij}(t).
    \end{split}
\end{equation}
 $h^{\textrm{quench}}_{ij}(t)$ is the quench Hamiltonian that drives the system from equilibrium.  We study two types of quenches here, written explicitly below as
\begin{equation}\label{quench}
    \begin{split}
        h^{\textrm{quench}}_{ij}(t) &= q\delta_{ij}f_\tau(t-t_0),\\
        h^{\textrm{quench}}_{ij}(t) &= q\delta_{ij}f_\tau(t-t_0)(1-f_\tau(t-t_1)),
    \end{split}
\end{equation}
where $f_\tau(t-t_0)$ is the Fermi-Dirac function,
\begin{equation}\label{fermi_dirac}
    f_\tau(t-t_0) = \frac{1}{1+e^{-\frac{t-t_0}{\tau}}}.
\end{equation}
It should also be understood that the chosen quench acts only on a specific subset of the chain ($N_q)$.  
\subsection{The Hartree-Fock Generalized\\Kadanoff-Baym Ansatz}
The HF-GKBA is an approximate partial solution to the full propagation of NEGF through the KBE.  The full Kadanoff-Baym equations are a set of five integro-differential equations.  In this paper we only consider the zero temperature limit, in which case the KBEs are given by\cite{Stan_2009}
\begin{equation}\label{KBE}
           \begin{split}
                i\partial_t G^{</>}(t,t') &= h^{\textrm{HF}}(t)G^{</>}(t,t') + I_1^{</>}(t,t')\\
                -i\partial_{t'} G^{</>}(t,t') &= G^{</>}(t,t')h^{\textrm{HF}}(t') + I_2^{</>}(t,t')\\
                i\partial_t G^{<}(t,t) &= [h^{\textrm{HF}}(t),G^{<}(t,t)] + I_1^<(t,t) - I_2^{<}(t,t)
            \end{split}
        \end{equation}
with 
\begin{equation}\label{coll_int}
    \begin{split}
        I_{1}^{</>}(t,t') &= \int_{0}^t \mathrm{d}\bar{t} \Sigma^R(t,\Bar{t})G^{</>}(\Bar{t},t') \\ &\hspace{20mm}+\int_{0}^{t'} \mathrm{d}\bar{t} \Sigma^{</>}(t,\Bar{t})G^{A}(\Bar{t},t')\\
        I_{2}^{</>}(t,t') &= \int_{0}^t \mathrm{d}\bar{t} G^R(t,\Bar{t})\Sigma^{</>}(\Bar{t},t') \\&\hspace{20mm}+ \int_{0}^{t'} \mathrm{d}\bar{t} G^{</>}(t,\Bar{t})\Sigma^{A}(\Bar{t},t').
    \end{split}
\end{equation}
Here $G^<(t,t')(G^>(t,t'))$ is the two time particle(hole) propagator.  The collision integrals, $I^{</>}_{1,2}$, take into account many-body correlation effects as well as system memory.  The two-time nature of the KBE combined with these integral terms leads to the cubic scaling of KBE mentioned in the introduction.  The HF-GKBA is derived directly from the KBE and can be summarized in the following equations\cite{Hermanns_2012},
\begin{equation}\label{HF-GKBA}
   \begin{split}
        G^{</>}(t,t') &= G^R(t,t')G^{</>}(t,t) - G^{</>}(t,t)G^R(t,t'),\\
        G^{R,A}&=\pm i \Theta(t_1 \pm t_2)T\{\mathrm{e}^{-i\int h^{\textrm{HF}}(t) dt}\}.
   \end{split}
\end{equation}
In other words, at each time step only the final equation in equation \eqref{KBE} is explicitly evaluated. Equation \eqref{HF-GKBA} is then used to reconstruct the time off-diagonal components.

Apart from those approximations made to the self-energy, which HF-GKBA and KBE share,  two additional approximations are made in the derivation of HF-GKBA.  The first involves neglecting certain integrals, similar to those in equation \eqref{coll_int}, over products of different components of the Green's function and self-energy.  These terms appear in the expression for reconstructing $G^{</>}(t,t')$ and are dropped, leading to the generalized Kadanoff-Baym ansatz(GKBA)\cite{PhysRevB.34.6933}. The HF-GKBA involves a further approximation where the full $G^{R/A}(t,t')$ are replaced by the retarded and advanced Hartree-Fock propagator.   The HF-GKBA still leave important quantities such as energy and particle number conserved as well as retaining causal time evolution.

Recently a linear time scaling($\sim O(N_t))$ implementation of the HF-GKBA has been achieved, opening the door for long-time evolution's of NEGFs\cite{PhysRevB.101.245101}.   The method removes the explicit appearance of integrals in equation \eqref{coll_int} from the differential equation for $G^<(t)$ by explicitly expressing them in terms of the correlated part of the equal time two-particle Green's function $\mathcal{G}(t)$. Within this formulation $\mathcal{G}(t)$ is propagated simultaneously with $G^<(t)$ using an equation analogous to the last line of equation \eqref{KBE}.  Throughout this paper, we use this propagation scheme to generate HF-GKBA results for the models discussed in section \ref{model_systems}.

The exact equation of motion for $\mathcal{G}(t)$ depends on the self-energy approximation used.  Throughout this paper, we use the $GW$ self-energy, due to its wide usage and its success in equilibrium condensed matter systems\cite{10.3389/fchem.2019.00377}.  For the $GW$ self-energy the equations of motion for $G^<(t)$ and $\mathcal{G}(t)$ in the orbital basis are given below.

\begin{equation}\label{G1-G2}
\begin{split}
       i \partial_t G^<_{ij}(t) &= [h^{\textrm{HF}}(t), G^<(t)]_{ij} + [I+I^\dagger]_{ij}(t)\\   i\partial_t \mathcal{G}_{ijkl}(t) &= [h^{(2),\textrm{HF}}(t),\mathcal{G}(t)]_{ijkl}\\&\hspace{13mm}+\Psi_{ijkl}(t) + \Pi_{ijkl}(t) - \Pi_{lkji}^*(t).
\end{split} 
\end{equation}
Above, the following definitions are made,
\begin{equation}
    \begin{split}
        h_{ij}^{\textrm{HF}}(t) &= h^{(0)}_{ij}(t) - i\sum_{kl} (w_{ikjl} - w_{iklj})(t)G_{kl}^<(t),\\
        I_{ij}(t)&=-i\sum_{klp} w_{iklp}(t)\mathcal{G}_{lpjk}(t),\\
        h^{(2),\textrm{HF}}_{ijkl}(t) &= \delta_{jl}h^{\textrm{HF}}_{ik}(t) + \delta_{ik}h^{\textrm{HF}}_{jl}(t),\\
        \Psi_{ijkl} &=\sum_{pqrs}w_{pqrs} \bigg[G^>_{ip}(t) G^<_{rk}(t) G^>_{jq}(t)G^<_{sl}(t)\\&\hspace{24mm}- G^<_{ip}(t)G^>_{rk}(t)G^<_{jq}(t)G^>_{sl}(t)\bigg],\\
    \end{split}
\end{equation}    \begin{equation*}
    \begin{split}     \Pi_{ijkl}&=\sum_{pqrs}w_{sqrp}(t)\bigg{[}G^>_{js}(t)G^<_{rl}(t)\\&\hspace{30mm}- G^<_{js}(t)G^>_{rl}(t)\bigg{]}\mathcal{G}_{ipkq}(t).
    \end{split}
\end{equation*}
Here $\Pi_{ijkl}$ accounts for polarization in the system and $\Psi_{ijkl}$ accounts for pair correlations built up due to two-particle scattering events\cite{PhysRevB.101.245101}.
\subsection{DMD}\label{sub_sect:DMD}
DMD
is a data-driven dimension reduction technique used to predict observables of a nonlinear dynamical system with a large number of degrees of freedom by constructing a low dimensional linear
dynamical model\cite{DMD0,schmid2011applications,kutz2016dynamic,TuRowley}. The linear model 
can be characterized by a number of spatial and temporal modes that can be obtained from the eigenvalues and eigenvectors of a linear operator known as a projected Koopman operator\cite{DMDtoKoop}.

To introduce the basic ideas of DMD, let us view~\eqref{G1-G2} as a general dynamical system of the form
\begin{equation}\label{eq:model}
    \frac{d\mathbf{x}(t)}{dt} = \mathbf{f}(\bx(t), t), \quad t\geq 0,
\end{equation}
where 
\begin{equation}
    \bx(t) = [\mathbf{g}_1(t), \mathbf{g}_2(t), ..., \mathbf{g}_n(t)]^T,
\end{equation}
and 
\begin{equation}
    \mathbf{g}_i(t) = [G^<_{i1}(t), G^<_{i2}(t), ..., G^<_{in}(t)]. \; 
\end{equation}
Here, $n$ is the number of sites, and
we simply consider the right-hand-side of\eqref{G1-G2} as a nonlinear function $\mathbf{f}: \mathbb{C}^n\otimes \mathbb{R}^+ \rightarrow \mathbb{C}^n$ of $\mathbf{x}$ and $t$.

The DMD method allows us to approximate \eqref{eq:model} by a linear model
\begin{equation}
    \frac{d\bx(t)}{dt} = \mathbf{A}\bx(t),
\end{equation}
with a carefully constructed operator $\mathbf{A}$.
For problems that have 
an explicit analytical expression of $\mathbf{f}(\mathbf{x}(t), t)$, it may be possible to linearize $\mathbf{f}(\mathbf{x}(t), t)$ and derive $\mathbf{A}$ explicitly. This linearization process essentially amounts to a linear response analysis.  However, when the analytical form of  $\mathbf{f}(\mathbf{x}(t), t)$ is unknown, performing such an analysis is difficult, if not impossible.

The linearization produced by DMD is based on the Koopman operator theory\cite{DMDtoKoop,Koopman1,Koopman2}, which is developed to characterize the evolution of a scalar observable function of $\mathbf{x}(t)$, denoted by $g(\mathbf{x}(t))$, to $g(\mathbf{x}(t+\Delta t))$ with $\Delta t>0$, i.e.
\[
g(\mathbf{x}(t+\Delta t)) = \mathcal{K}_{\Delta t} g(\mathbf{x}(t)).
\]

In the limit of $\Delta t \rightarrow 0$, the Koopman operator defines a linear dynamical system
\begin{equation}\label{eq:Koopman}
\frac{dg(\mathbf{x}(t))}{dt} = \mathcal{K} g(\mathbf{x}(t)).
\end{equation}

Because the Koopman operator $\mathcal{K}$ is a linear operator that maps from a function space to 
another function space, it has an infinite number of eigenvalues $\lambda_j$ and eigenfunctions 
$\varphi_j(\mathbf{x})$, $j = 1, 2, ..., \infty$. 

If the observable functions of interest form an invariant subspace of 
$\mathcal{K}$ spanned by a finite subset of eigenvalues and eigenvectors, then it is possible to 
construct a finite-dimensional operator (matrix) approximation to $\mathcal{K}$.

To be specific, if $g_1(\mathbf{x})$, $g_2(\mathbf{x})$,...,$g_n(\mathbf{x})$ 
are $n$ observable functions so that

\begin{equation}\label{eq:KoopmanA}
\begin{bmatrix}
g_1(\mathbf{x}) \\
g_2(\mathbf{x}) \\
\vdots \\
g_n(\mathbf{x}) 
\end{bmatrix}
 = 
\mathbf{V}_1
\begin{bmatrix}
\varphi_1(\mathbf{x}) \\
\varphi_2(\mathbf{x}) \\ 
\vdots \\
\varphi_k(\mathbf{x})
\end{bmatrix}
 = 
\mathbf{V}_1\mathbf{V}_2
\begin{bmatrix}
g_1(\mathbf{x}) \\
g_2(\mathbf{x}) \\ 
\vdots \\
g_k(\mathbf{x})
\end{bmatrix}
\end{equation}

for some $k\in\mathbb{N}$ and matrices $\mathbf{V}_1 \in \mathbb{C}^{n\times k}$, $\mathbf{V}_2 \in\mathbb{C}^{k\times n}$, 
then $\mathcal{K}$ can be approximated by a $n \times n$ matrix $\mathbf{A} = \mathbf{V}_1\mathbf{V}_2$ on these observable functions.

But in practice, we cannot assume that equation \eqref{eq:KoopmanA} holds for our observable functions, so we can only get a finite-dimension approximation of $\mathcal{K}$ represented by matrix $\mathbf{A}$. To construct such an approximation, observable functions are chosen to be the components of
$\mathbf{x}(t)$ defined in equation \eqref{eq:model}, we take snapshots of $\mathbf{x}(t)$ at $t_j = (j-1)\Delta t$, i.e., $\mathbf{x}_j = \mathbf{x}(t_j)$, for $j=1,...,m$, and use them to build two matrices 
$\mathbf{X}_1$ and $\mathbf{X}_2$ of the form
\begin{equation}
\mathbf{X}_1 =\left( \mathbf{x}_1 \: \mathbf{x}_2 \: \cdots \: \mathbf{x}_{m-1} \right) \ \ \mbox{and} \ \
\mathbf{X}_2 =\left( \mathbf{x}_2 \: \mathbf{x}_3 \: \cdots \: \mathbf{x}_{m} \right).
\label{eq:mats}
\end{equation}
The creation of these matrices requires the explicit time propagation of the equations of motion up to time $t_m$. The finite-dimensional approximation to the Koopman operator can then be obtained by solving the following linear least squares problem
\begin{equation}
 \min_{\mathbf{A}} \| \mathbf{A} \mathbf{X}_1 - \mathbf{X}_2 \|_F^2.
\label{eq:lsq}
\end{equation}
The solution to \eqref{eq:lsq}  is
\begin{equation}\label{eq:A}
    \mathbf{A} = \mathbf{X}_2\mathbf{X}_1^\dagger,
\end{equation}
where $\mathbf{X}_1^\dagger$ is the Moore-Penrose pseudoinverse of $\mathbf{X}_1$ that can be computed from the singular value decomposition (SVD)\cite{SVD} of $\mathbf{X}_1$.  Once $\mathbf{A}$ is calculated using the $m$ snapshots of $\mathbf{x}(t)$, it can be used to further approximately evolve the system for times $t_j$ for $j>m$. In essence the $\mathbf{A}$ computed from equation \eqref{eq:A} is used as a generator of the time evolution for times after $t_m$.
If the nonzero singular values of $\mathbf{X}_1$, $\sigma_j$, $j = 1,2,...,m$, decrease rapidly with respect to $j$, which indicates that the numerical rank, denoted by $r$, of $\mathbf{X}_1$ is much smaller than $m$ and $n$, we can  use a truncated SVD of $\mathbf{X}_1$ in the form of $\mathbf{X}_1 = \widetilde{\mathbf{U}}\widetilde{\mathbf{\Sigma}}\widetilde{\mathbf{V}}^T$, where the $r\times r$ diagonal matrix $\widetilde{\mathbf{\Sigma}}$ contains the leading $r$ dominant  singular values of $\mathbf{X}_1$, and $\widetilde{\mathbf{U}}$ and $\widetilde{\mathbf{V}}$ 
contain the corresponding right and left singular vectors, to obtain an approximation of $\mathbf{A}$ as
\begin{equation}
\mathbf{A} \approx \mathbf{X}_2\widetilde{\mathbf{V}}\widetilde{\mathbf{\Sigma}}^{-1}\widetilde{\mathbf{U}}^*.
\label{eq:projA}
\end{equation}

We can now fully characterize the approximated reduced order linear dynamical system model by 
diagonalizing the projected Koopman operator
$\widetilde{\mathbf{A}} = \widetilde{\mathbf{U}}^\ast \mathbf{A} \widetilde{\mathbf{U}} 
= \widetilde{\mathbf{U}} \mathbf{X}_2 \widetilde{\mathbf{V}} \widetilde{\mathbf{\Sigma}}^{-1}\in\mathbb{C}^{r\times r}$.
Let 
\begin{equation}
\widetilde{\mathbf{A}}\mathbf{W} = \mathbf{W}\mathbf{\Lambda}
\label{eq:dmdev}
\end{equation}
be the eigendecomposition of $\widetilde{\mathbf{A}}$, 
where $\mathbf{\Lambda} = {\rm{diag}}(\lambda_1, ..., \lambda_r)$ is composed of the eigenvalues of $\widetilde{\mathbf{A}}$, and the columns of $\mathbf{W}$ are the corresponding eigenvectors.
The matrix
\begin{equation}
\mathbf{\Phi} = \mathbf{X}_2\widetilde{\mathbf{V}}\widetilde{\mathbf{\Sigma}}^{-1}\mathbf{W}
\label{eq:dmdmodes}
\end{equation}
contains the so-called DMD modes. If $\phi_{\ell}$ is the $\ell$th column of $\mathbf{\Phi}$, the DMD approximation to $\mathbf{x}$ can be represented by
\begin{equation}\label{eq:evol_DMD}
\mathbf{x}(t)\approx \sum_{\ell=1}^r\mathbf{\phi}_\ell\exp(i\omega_\ell^{\text{DMD}} t)b_\ell = \mathbf{\Phi}\exp(\mathbf{\Omega} t)\mathbf{b}.
\end{equation}
where $\omega_\ell^{\text{DMD}} = -i{\ln{\lambda_\ell}}/{\Delta t}$, $\ell = 1, ..., r$, $\mathbf{\Omega} = {\ln{\mathbf{\Lambda}}}/{\Delta t} = {\rm{diag}}(i\omega_1^{\rm{DMD}}, ..., i\omega_r^{\rm{DMD}})$, and the amplitude vector $\mathbf{b}:=[b_1, ..., b_r]^T$ is taken either as the projection of the initial value $\mathbf{x}_1$ onto the DMD modes, i.e.,
\begin{equation}\label{eq:b1}
\mathbf{b} = \mathbf{\Phi}^\dagger \mathbf{x}_1,
\end{equation}
or as the least squares fit of equation \eqref{eq:evol_DMD} on the sampled trajectories, i.e.,
\begin{equation}\label{eq:b2}
\mathbf{b} = \arg\min_{\tilde{\mathbf{b}}\in\mathbb{C}^n}\sum_{j=1}^m\|\mathbf{\Phi}\exp(\mathbf{\Omega} t_j)\tilde{\mathbf{b}}-\bx_j\|^2,
\end{equation}
where $\|\cdot\|$ denotes the standard Euclidean norm of a vector. For more details on the numerical procedure, we refer readers to references\cite{DMD0,kutz2016dynamic,TuRowley,DMDdiag,DMDtwotime}.

The major computational cost of DMD computation is in the SVD of $\mathbf{X}_1$, which is $O(\min(m^2n, mn^2))$. The memory cost is $O(mn)$.

\section{Methods}\label{sec:Methods}
 For systems with up to 8 sites at half-filling we prepare trajectories for the model and quenches described in section \ref{Theory} using both exact diagonalization and HF-GKBA. A publicly available version of the code used in these simulations is available online\footnote{https://github.com/VlcekGroup/G12KBA.git}. We also use HF-GKBA to create a trajectory for 16 sites, however exact time evolution was not possible for this system size.  Equation \eqref{G1-G2} was propagated using fourth-order Runge-Kutte with a time step of $0.07 J^{-1}$, and we use the same time step for the exact diagonalization propagation.  We performed calculations for two values of the decay parameter $\gamma$, firstly we take $\gamma=\infty$ which leads to a model with onsite interactions only.  The second case we study has $\gamma = 0.7$ so that the nearest neighbour is subject to approximately half the interaction of the onsite interaction strength.  For each model and quench, calculations were run for $U = 0.1J$, $0.3J$, $0.5J$ and $1.0J$.  Additionally, we explored the effect of quenching different portions of the system, testing both half and quarter system quenches.

For both HF-GKBA and exact diagonalization we prepare the system in the respective correlated initial state before initiating the quench.   In the case of exact diagonalization, we can trivially prepare the system in the exact ground state, by diagonalizing and finding the eigenstate with the lowest energy.  For the HF-GKBA we start by preparing the system in the non-interacting ground state and then time evolves with equation \eqref{G1-G2} while slowly turning on the $U$ parameter.  We choose the Fermi-Dirac function introduced in equation \eqref{fermi_dirac} as our switching function.
We found the values $t_0 = 25 $ and $\tau = 3$ gave a sufficiently slow rate of switching to converge the models and parameters presented here. 

For the alternating step potential in equation \eqref{single_particle} we chose a value of $V=2$.  We found for $N_s\leq16$, this value of $V$ opened the gap sufficiently to perform the adiabatic switching procedure successfully. However, we note that for $V = 2$ the adiabatic switching procedure became unstable as we went to larger systems($N_s=32)$.
\section{Results}
In total, around 100 different system setups were tested, see section \ref{sec:Methods}, in this section we will present a small representative selection of these results.   In section \ref{onsite_vs_lr}, we look at the performance HF-GKBA for the long-range and onsite models.  We compare trajectories for different values of $U$ and for the two quenches described in section \ref{model_systems}.  In section \ref{DMD_results} we show results demonstrating the ability of the DMD to fit the Green's function in the long-range model with 16 sites.

\begin{figure*}
    \centering
    \includegraphics[width=\linewidth]{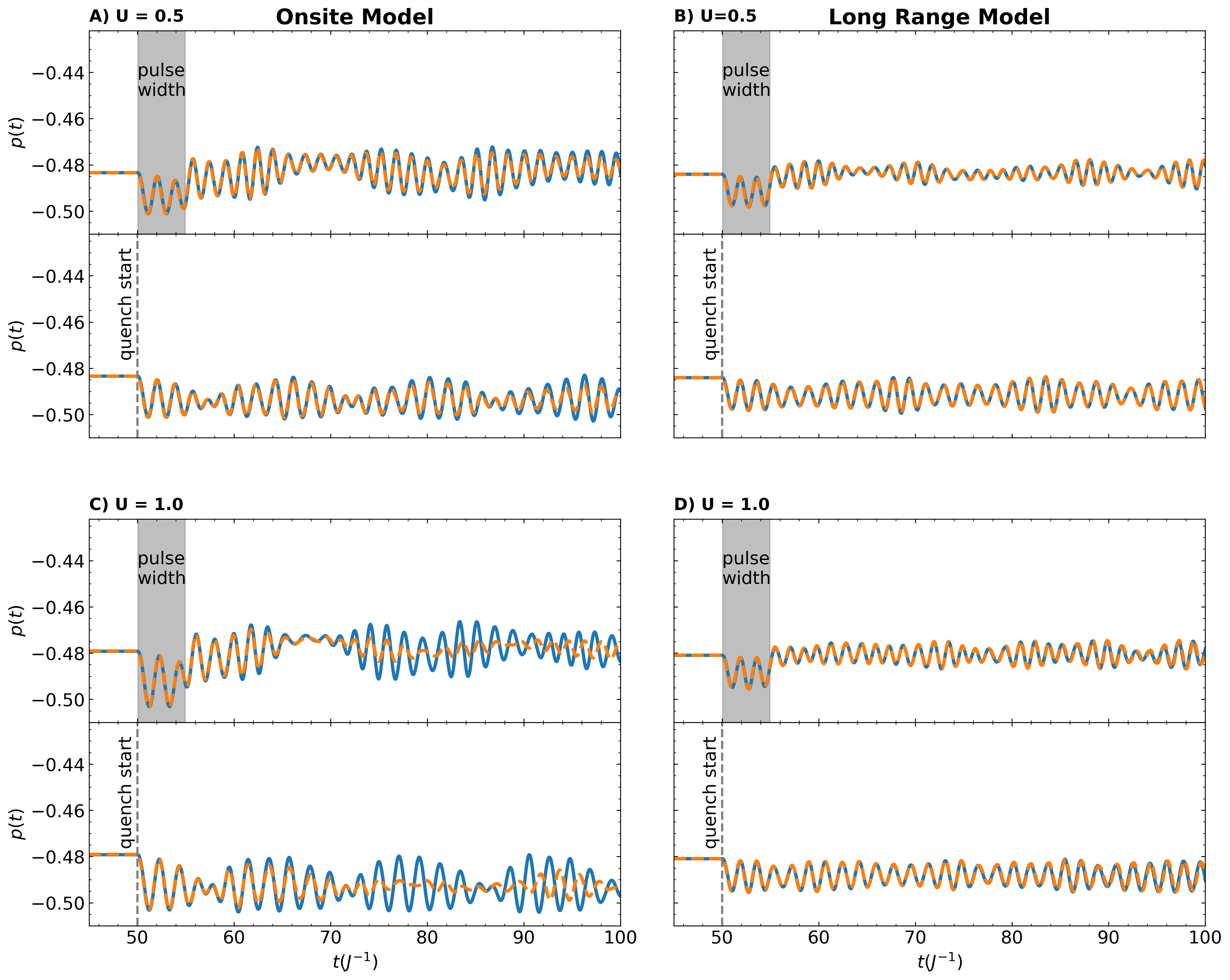}
    \caption{Comparison of the early time dynamics of the dipole given by HF-GKBA(orange-dashed) to exact results(blue-solid) for $\gamma = \infty$(left) and $\gamma = 0.7$(right) with $U = 0.5J$ and $U = 1.0J$ and for $N_s=8$. Figures 1 (2) of each panel A,B,C and D show the dipole after the first half the system is pulse quenched (fully quenched).  All the above figures are plotted on the same scale.}
    \label{EHM_vs_HM}
\end{figure*}

\subsection{HF-GKBA Dynamics for Onsite  and Long Range Models}\label{onsite_vs_lr}
First, we analyze the early time dynamics of both models and compare the trajectories to exact diagonalization.  In Fig. \ref{EHM_vs_HM} we show the dipole generated by HF-GKBA and the exact diagonalization for 8 sites for each of the Hamiltonians described by equations \eqref{two-body} and \eqref{single_particle}. The Green's function contains $N_s^2$ elements and so to compress this large quantity of time trajectories we choose the center of mass dipole as our figure of comparison, which we calculate as
\begin{equation}
    p(t) = \frac{i}{N_s}\sum_{j=1}^{N_s}\left(\frac{N_s-1}{2} - j\right)[G^<_{jj}(t) - G^<_{N_s-j+1N_s-j+1}(t)],
\end{equation}
where $G^<_{jj}(t)$ is the density on site $j$ at time $t$.
We choose the dipole for two reasons, besides compressing the Green's function to a single number. Firstly, we believe it is most relevant to test the ability of the HF-GKBA to reproduce experimental observables. Secondly, the dipole is an integrated quantity and so provides a site-independent accumulated result of the density, which also removes some of the bias due to edge effects or the quench location.  We found the dipole to be a representative quantity for the results in the following sections.
\FloatBarrier
\begin{figure*}
    \centering
    \includegraphics[width = \textwidth]{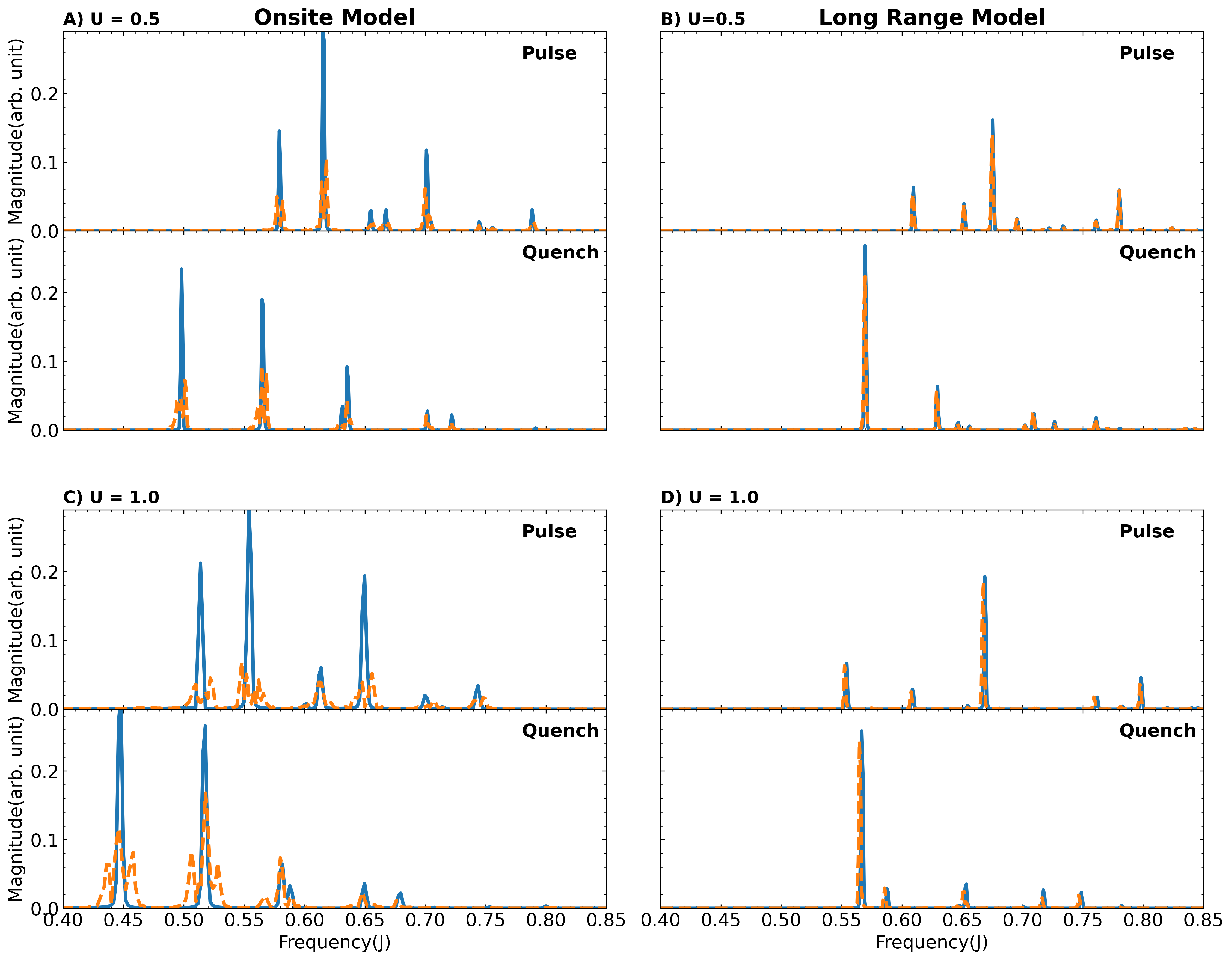}
    \caption{Comparison of frequency spectrum of the dipole for HF-GKBA(orange-dotted) and the exact result(blue-solid) for each of the cases in Fig. \ref{EHM_vs_HM}. Figures 1 (2) of each panel A,B,C and D show the dipole frequency spectrum after the first half of the system is pulse quenched (fully quenched). All figures are plotted on the same scale. We also note that for sub-panel C, due to numerical instability only 500 time units of the exact and HF-GKBA trajectory were used to create the spectrum.}
    \label{fig:freq_spectrum}
\end{figure*}

\FloatBarrier
The results in Fig. \ref{EHM_vs_HM} show the first $50$ time units after the quench at $t=50$, and we direct the reader to the SI for the full dipole trajectory over 1000 time units. After preparing the system in the correlated initial state we perform a full or pulsed quench of magnitude 1 to the first 4 sites.  Quenches were also tested on the first 2 sites, however, little qualitative difference was observed between the two cases.  For the full quench, the parameters chosen were $t_0 = 50$ and $\tau = 0.2$.   In the case of the pulse quench, we fix $t_0$ and $\tau$ to be the same as for the full quench and take $t_1 = 55$.  We note that for $U=1.0$ the HF-GKBA trajectory for the onsite model becomes unstable and diverges after between 500 and 800 time units depending on the quench type, whereas the extended model remains stable for the entire trajectory.
From the portion of the trajectory shown in Fig. \ref{EHM_vs_HM}, we see that for $U = 0.5J$ the HF-GKBA captures the dynamics of the onsite model and long-range model quite well.  However, already at around 40 time units after the pulse we see the amplitude of the HF-GKBA result begin to decrease relative to the exact. Meanwhile the long-range model continues to match the dynamics remarkably well for the entire trajectory shown.  As we move to $U = 1.0J$ difference between the HF-GKBA in the onsite and long-ranged model becomes even more extreme.  At around $20$ time units after the pulse the HF-GKBA fails completely to capture the true dipole dynamics, whereas in the long-range model it is matched almost exactly by the HF-GKBA.

To compare the HF-GKBA and exact results over the full dipole trajectory we look at the frequency spectrum for each of the trajectories shown in Fig. \ref{EHM_vs_HM}.

 Due to the failure of HF-GKBA in the onsite model for $U=1.0$ we only use the first 500 time units to generate the frequency spectrum.
  However, for the remaining results, we used the entire trajectory from the time after the quench. We found the frequency spectrum to be a more reliable measure of quality than the residual between HF-GKBA and the exact result.  In particular, in the long-range model, the primary error that arose was a phase mismatch between the HF-GKBA and exact.  In this case, the residual provides a misleading measure of the performance of HF-GKBA.  In Fig. \ref{fig:freq_spectrum} we only show the frequency range $0.40J$ to $0.85 J$ as this range held the major spectrum peaks. The full spectrum is included in Fig. S2 of the SI.

For the onsite model, we see quite good agreement in the peak positions between the exact and HF-GKBA results.  Clearly, the HF-GKBA consistently underestimates the magnitude of the spectrum peaks, and for the full quench the HF-GKBA incorrectly identifies the maximum frequency peak. As we go from $U = 0.5J$ to $U = 1.0J$ we see a broadening of the peaks both in exact and HF-GKBA results, which is partially related to the shorter trajectory used to create the spectrum. The HF-GKBA does however overestimate the broadening and even leads to the formation of additional peaks in the spectrum. 
The HF-GKBA continues to capture the peak positions well in the long-range model. Furthermore, we now observe that the amplitudes of each frequency peak match the exact result far better than in the onsite model.  Going to $U = 1.0J$ the amplitudes and peak positions continue to be matched very well. We point out there is a slight shift in the HF-GKBA peaks $U=1.0J$, this causes the dipole trajectories to move slowly in and out of phase with one another over the time evolution.  
From these results, in the scenarios we have studied so far, we see a clear improvement of the HF-GKBA upon the inclusion of exponentially decaying interactions in the model.  The improvement holds over relatively long times leading to a spectrum that matches the exact spectrum almost perfectly. We also note a similar improvement in the case of a $\frac{1}{r}$ decay, which is shown in Fig. S6 of the supplementary for a 4-site model.  

In the following section, we will look at results obtained from applying DMD to fit the HF-GKBA Green's function.

\begin{figure*}
    \centering
    \includegraphics[width=\textwidth]{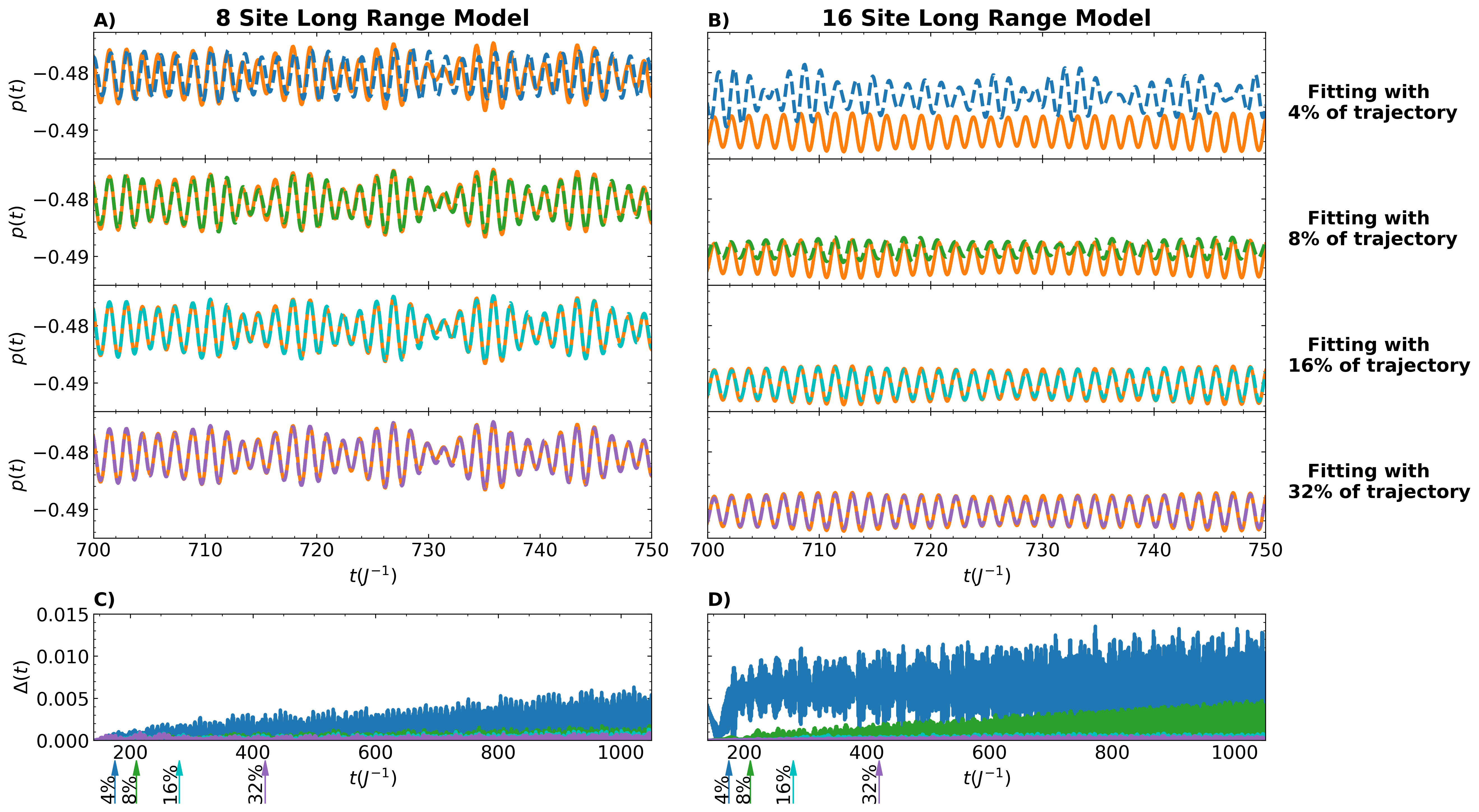}
    \caption{DMD Trajectories(dashed-lines) for long-range models with 8 and 16 sites for different sizes of the fitting window with a pulse quench on half the sites compared to HF-GKBA(solid-orange).  Sub-panels A and B show the dipole between 700-750 time units for HF-GKBA and the DMD reconstruction.  Sub-panels C and D show the residual between DMD and HF-GKBA dipole for each size fitting window.}
    \label{fig:DMD_fit}
\end{figure*}
\subsection{DMD Extrapolation of HF-GKBA Trajectories}\label{DMD_results}
In the previous section, we observed excellent agreement between the HF-GKBA and the exact time evolution for the long-range model.  This motivates the use of DMD in conjunction with HF-GKBA to extrapolate long time trajectories from partial trajectories of HF-GKBA Green's functions. In this section, we present results investigating the effectiveness of DMD in predicting the dynamics of the HF-GKBA. We use the DMD procedure outlined in section \ref{sub_sect:DMD} and apply it to various portions of the total Green's function trajectory. 
Relative to the long-range model we found for the onsite model a much larger portion of the trajectory was needed to produce reasonable results, and for most cases, DMD did not produce a successful reconstruction of the Green's function trajectory.  Because of the poor performance of DMD as well as the poor performance of the HF-GKBA for the onsite model we omit these results and instead discuss possible reasons for the failure in section \ref{sec:discussion}.

In Fig. \ref{fig:DMD_fit} we show results for the DMD extrapolation of $G^<(t)$ for the long-range model with 8 and 16 sites for $U = 1.0J$ and a pulse quench on half of the sites.  For the smaller models we tested similar behaviour was found, so we omit these results. Panels A and B of Fig. \ref{fig:DMD_fit} show the dipole for 4 different sized fitting windows.  For each window, DMD is used to construct a reduced-order model for the total Green's function.  The reduced order model is then used to extrapolate beyond the fitting window. We again choose to report the dipole of the system for the same reasons listed in section \ref{onsite_vs_lr}.

The DMD dipole is generated by using snapshots of the HF-GKBA Green's function to construct the DMD extrapolation model.  This extrapolated result is then used to calculate the dipole.  In the bottom two panels of Fig. \ref{DMD_results} the residuals between the DMD extrapolated dipole and the HF-GKBA dipole over the full trajectory for each size of the fitting window are shown.

We found the ability of DMD to reconstruct the Green's function was captured well by comparing the DMD dipole to the HF-GKBA dipole.  Similar plots to that shown in Fig. \ref{fig:DMD_fit} are shown  in Fig. S3, S4 and S5 of the SI for a selection of components of the Green's function.  As expected, we see the residual between the DMD extrapolated dipole and the HF-GKBA dipole decrease as the fitting window is increased. These figures suggest that for the 8-site long-range model, somewhere between 4\% and 8\% of the total trajectory is needed to have a good fit of $G^{<}(t)$, and this fraction increases to between 8\% and 16\% when we go to the 16 site model.  This is likely due to more low-frequency modes being present in the 16-site model, which requires a longer fitting window for DMD.  In section \ref{sec:discussion} the implications and prospects of DMD will be discussed.

\section{Discussion and Conclusions}\label{sec:discussion}
Returning to the results comparing the exact and HF-GKBA propagation, we first discuss the significant improvement of HF-GKBA upon going to the long-range model.  We suggest two possible contributions to this observation.  Firstly, we note the magnitude of oscillations is smaller for the extended model.  We explain this by noting the higher degree of repulsive couplings between sites increases the localization of individual particles. This may in turn lead to dynamics that are easier to capture with HF-GKBA.  A second possibility comes from the self-energy approximation used in these calculations. In systems where screening is important the $GW$ self-energy becomes the dominant contribution to the full self-energy and describes well the many-body interactions between particles. In the onsite model due to the completely local interactions the amount of screening will be quite small and so the $GW$ self-energy will not capture the physics well. However, for a more realistic setup with long-range interactions (encountered in most materials to which GW is meaningfully applied), we see that the HF-GKBA behavior is significantly closer to the ED results.  We also see a similar behaviour for the case of a long-range interaction with $\frac{1}{r}$ decay, see Fig. S6 in the SI.  We will investigate further the limitations of the various self-energy formulations, as recent works proposed a route to construct reliable higher order (i.e., beyond $GW$) schemes that help with the description of excited states in equilibrium\cite{PhysRevB.106.165129}.  

We will investigate further the limitations of the various self-energy formulations. A recent work has shown good agreement between exact results and HF-GKBA using the T-matrix self energy for the onsite Hubbard model up to $U=4.0$\cite{PhysRevB.105.125135}. 
Furthermore, a recent work proposed a route to construct reliable higher order (i.e., beyond $GW$) schemes that include T-matrix corrections on top of $GW$ and help with the description of excited states in equilibrium\cite{PhysRevB.106.165129}. Incidentally, in equilibrium, the inclusion of these T-matrix correction terms also extends the validity of the approximation up to $U=4.0$ in the Hubbard dimer.  Thus we assume the agreement between the HF-GKBA and exact diagonalization results in this paper can be improved for all models studied here by changing the self-energy or by including additional correction terms.

We believe the poor performance of DMD in onsite model can be explained at least partially if we look at the full frequency spectrum for the dipole shown in Fig. S2 of the SI. Clearly, the onsite model has a much larger low-frequency component than the long-range models.  Lower frequencies are more difficult to capture using DMD since a longer portion of the trajectory needs to be sampled to observe these long time modes.  We believe this to be part of the reason why DMD tends to fail in the onsite model.  We also point this out as one of the limitations of DMD, since for systems with very low-frequency modes DMD will have to be performed from a very large portion of the trajectory.   Similarly, in the case of the 8 and 16 site models a downward shift in the low-frequency part of the spectrum going from $N_s=8$ to $N_s = 16$, which at least partially explains why a larger amount of the trajectory was required by DMD

We found that for the long-time trajectories prepared in this work, several exhibited numerical instabilities after several hundred time units.  These errors typically arose for $U \geq 1.0$ and became more prominent as we moved to larger systems.  We comment on two possible sources for these errors. The first possible way these errors arise is described in detail in \cite{PhysRevB.105.165155} and is related to certain consistency relations for the two-particle Green's function breaking down at long times and strong couplings. We point out that the procedure for enforcing the consistency relations function is extremely expensive as it requires the diagonalization of the two-particle Green's function throughout the time-stepping procedure, which scales as $O(N^6)$.  If not corrected, these inconsistencies can lead to divergences of the HF-GKBA solution.  The second possible source of error may arise from the adiabatic switching procedure.   It is possible that for some of the parameters and models we tested small residual errors from the adiabatic switching preparation built up and contributed to the failures of the HF-GKBA time propagation.  

These issues offer another opportunity for the use of DMD to assist in the propagation of NEGFs. If the trajectory fails after a sufficiently long time and one can clearly identify the point of failure, then DMD can be used to fit $G^{<}(t)$ on the portion of the trajectory before the solution breaks down.   The DMD fitted result can be propagated in place of explicit propagation of equations \eqref{G1-G2}.  

In this paper, we have presented results demonstrating a vast improvement of the HF-GKBA when long-range interactions are included.  We also observed that DMD is a suitable tool for the reconstruction of long-time trajectories of the HF-GKBA Despite this we still believe DMD can be a powerful tool to be used alongside HF-GKBA and the G1-G2 scheme, especially in the long time propagation of Green's functions in large systems.  In future work, we will continue to explore DMD as a way of preparing trajectories for large-scale systems.  We will also investigate generalising existing stochastic techniques used in equilibrium systems\cite{PhysRevB.98.075107,doi:10.1021/acs.jctc.7b00770,PhysRevLett.113.076402} to non-equilibrium.  Combining stochastic approaches in conjunction with DMD we hope push HF-GKBA to explore the physics of large multi-band systems.
\section*{Acknowledgements}
This material is based upon work supported by the U.S. Department of Energy, Office of Science, Office of Advanced Scientific Computing Research and Office of Basic Energy Sciences, Scientific Discovery through Advanced Computing (SciDAC) program under Award Number DE-SC0022198
\bibliographystyle{unsrt}
\bibliography{Bib}

\begin{thebibliography}{10}

\bibitem{RevModPhys.81.163}
M.~Krausz, F.and~Ivanov.
\newblock Attosecond physics.
\newblock {\em Rev. Mod. Phys.}, 81:163--234, Feb 2009.

\bibitem{Le_Hur_2016}
K.~Le Hur, L.~Henriet, A.~Petrescu, K.~Plekhanov, G.~Roux, and M.~Schir{\'{o}}.
\newblock Many-body quantum electrodynamics networks: Non-equilibrium condensed
  matter physics with light.
\newblock {\em Comptes Rendus Physique}, 17(8):808--835, oct 2016.

\bibitem{stefanucci2013nonequilibrium}
G.~Stefanucci and R.~van Leeuwen.
\newblock {\em Nonequilibrium Many-Body Theory of Quantum Systems: A Modern
  Introduction}.
\newblock Cambridge University Press, 2013.

\bibitem{Kadanoff}
L.P. Kadanoff and G.~Baym.
\newblock {\em {Quantum Statistical Mechanics}}.
\newblock W.A. Benjamin Inc., New York, 1962.

\bibitem{Stan_2009}
A.~Stan, N.E. Dahlen, and Robert van Leeuwen.
\newblock Time propagation of the kadanoff{\textendash}baym equations for
  inhomogeneous systems.
\newblock {\em The Journal of Chemical Physics}, 130(22):224101, jun 2009.

\bibitem{von_Friesen_2009}
M.~Puig von Friesen, C.~Verdozzi, and C.-O. Almbladh.
\newblock Successes and failures of kadanoff-baym dynamics in hubbard
  nanoclusters.
\newblock {\em Physical Review Letters}, 103(17), oct 2009.

\bibitem{Friesen_2010}
M.~Puig von Friesen, C.~Verdozzi, and C.-O. Almbladh.
\newblock Artificial damping in the kadanoff-baym dynamics of small hubbard
  chains.
\newblock {\em Journal of Physics: Conference Series}, 220(1):012016, apr 2010.

\bibitem{PhysRevB.90.125111}
S.~Hermanns, N.~Schl\"unzen, and M.~Bonitz.
\newblock Hubbard nanoclusters far from equilibrium.
\newblock {\em Phys. Rev. B}, 90:125111, Sep 2014.

\bibitem{bonitz2015quantum}
M.~Bonitz.
\newblock {\em {Quantum Kinetic Theory}}.
\newblock Springer International Publishing, 2015.

\bibitem{PhysRevB.34.6933}
P.~Lipavsk\'y, V.~\ifmmode \check{S}\else \v{S}\fi{}pi\ifmmode~\check{c}\else
  \v{c}\fi{}ka, and B.~Velick\'y.
\newblock {Generalized Kadanoff-Baym} ansatz for deriving quantum transport
  equations.
\newblock {\em Phys. Rev. B}, 34:6933--6942, Nov 1986.

\bibitem{Hermanns_2012}
S.~Hermanns, K.~Balzer, and M.~Bonitz.
\newblock {The non-equilibrium Green function approach to inhomogeneous quantum
  many-body systems using the generalized Kadanoff{\textendash}Baym ansatz}.
\newblock {\em Physica Scripta}, T151:014036, nov 2012.

\bibitem{PhysRevLett.128.016801}
E.~Perfetto, Y.~Pavlyukh, and G.~Stefanucci.
\newblock {Real-Time {$GW$} : Toward an ab initio description of the ultrafast
  carrier and exciton dynamics in two-dimensional materials}.
\newblock {\em Phys. Rev. Lett.}, 128:016801, Jan 2022.

\bibitem{PhysRevB.101.245101}
J.-P. Joost, N.~Schl\"unzen, and M.~Bonitz.
\newblock G1-{G}2 scheme: Dramatic acceleration of nonequilibrium green
  functions simulations within the hartree-fock generalized kadanoff-baym
  ansatz.
\newblock {\em Phys. Rev. B}, 101:245101, Jun 2020.

\bibitem{PhysRevB.105.165155}
J.-P. Joost, N.~Schl\"unzen, H.~Ohldag, M.~Bonitz, F.~Lackner, and
  I.~B\ifmmode~\check{r}\else \v{r}\fi{}ezinov\'a.
\newblock {Dynamically screened ladder approximation: Simultaneous treatment of
  strong electronic correlations and dynamical screening out of equilibrium}.
\newblock {\em Phys. Rev. B}, 105:165155, Apr 2022.

\bibitem{PhysRevLett.113.076402}
D.~Neuhauser, Y.~Gao, C.~Arntsen, Cyrus Karshenas, E.~Rabani, and R.~Baer.
\newblock Breaking the theoretical scaling limit for predicting quasiparticle
  energies: The stochastic {$GW$} approach.
\newblock {\em Phys. Rev. Lett.}, 113:076402, Aug 2014.

\bibitem{doi:10.1021/acs.jctc.7b00770}
V.~Vlček, E.~Rabani, D.~Neuhauser, and R.~Baer.
\newblock Stochastic {$GW$} calculations for molecules.
\newblock {\em Journal of Chemical Theory and Computation}, 13(10):4997--5003,
  2017.
\newblock PMID: 28876912.

\bibitem{PhysRevB.106.165129}
C.~Mejuto-Zaera and V.~Vl\ifmmode~\check{c}\else \v{c}\fi{}ek.
\newblock Self-consistency in {$GW\mathrm{\ensuremath{\Gamma}}$} formalism
  leading to quasiparticle-quasiparticle couplings.
\newblock {\em Phys. Rev. B}, 106:165129, Oct 2022.

\bibitem{10.3389/fchem.2019.00377}
D.~Golze, M.~Dvorak, and P.~Rinke.
\newblock The {GW} compendium: A practical guide to theoretical photoemission
  spectroscopy.
\newblock {\em Frontiers in Chemistry}, 7, 2019.

\bibitem{DMD0}
P.~J. Schmid.
\newblock Dynamic mode decomposition of numerical and experimental data.
\newblock {\em J. Fluid Mech.}, 656:5--28, 2010.

\bibitem{schmid2011applications}
P.~J. Schmid, L.~Li, M.~P. Juniper, and O.~Pust.
\newblock Applications of the dynamic mode decomposition.
\newblock {\em Theor. Comput. Fluid Dyn.}, 25(1):249--259, 2011.

\bibitem{kutz2016dynamic}
J.~N. Kutz, S.~L. Brunton, B.~W. Brunton, and J.~L. Proctor.
\newblock {\em Dynamic mode decomposition: data-driven modeling of complex
  systems}.
\newblock SIAM, 2016.

\bibitem{TuRowley}
J.~H. Tu, C.~W. Rowley, D.~M. Luchtenburg, S.~L. Brunton, and J.~N. Kutz.
\newblock On dynamic mode decomposition: Theory and applications.
\newblock {\em J. Comput. Dyn.}, 1(2):391--421, 2014.

\bibitem{DMDtoKoop}
H.~Arbabi and I.~Mezi\'{c}.
\newblock Ergodic theory, dynamic mode decomposition, and computation of
  spectral properties of the {Koopman} operator.
\newblock {\em SIAM J. Appl. Dyn. Syst.}, 16(4):2096--2126, 2017.

\bibitem{Koopman1}
B.~O. Koopman.
\newblock Hamiltonian systems and transformation in {Hilbert} space.
\newblock {\em Proc. Natl. Acad. Sci. U.S.A.}, 17(5):315, 1931.

\bibitem{Koopman2}
B.~O. Koopman and J.~von Neumann.
\newblock Dynamical systems of continuous spectra.
\newblock {\em Proc. Natl. Acad. Sci. U.S.A.}, 18(3):255, 1932.

\bibitem{SVD}
G.~H. Golub and C.~F. Van~Loan.
\newblock {\em Matrix computations}, volume~3.
\newblock JHU press, 2013.

\bibitem{DMDdiag}
J.~Yin, Y.-h. Chan, F.~da~Jornada, D.~Qiu, C.~Yang, and S.~G. Louie.
\newblock Analyzing and predicting non-equilibrium many-body dynamics via
  dynamic mode decomposition.
\newblock {\em arXiv preprint arXiv:2107.09635}, 2021.

\bibitem{DMDtwotime}
J.~Yin, Y.~h.~Chan, F.~H. da~Jornada, D.~Y. Qiu, S.~G. Louie, and C.~Yang.
\newblock Using dynamic mode decomposition to predict the dynamics of a
  two-time non-equilibrium {Green’s function}.
\newblock {\em J. Comput. Sci.}, 64:101843, 2022.

\bibitem{Note1}
https://github.com/VlcekGroup/G12KBA.git.

\bibitem{PhysRevB.105.125135}
Y.~Pavlyukh, E.~Perfetto, Daniel Karlsson, Robert van Leeuwen, and
  G.~Stefanucci.
\newblock Time-linear scaling nonequilibrium green's function method for
  real-time simulations of interacting electrons and bosons. ii. dynamics of
  polarons and doublons.
\newblock {\em Phys. Rev. B}, 105:125135, Mar 2022.

\bibitem{PhysRevB.98.075107}
V.~Vl\ifmmode\check{c}\else\v{c}\fi{}ek, W.~Li, R.~Baer, E.~Rabani, and
  D.~Neuhauser.
\newblock Swift {$GW$} beyond 10,000 electrons using sparse stochastic
  compression.
\newblock {\em Phys. Rev. B}, 98:075107, Aug 2018.

\end{thebibliography}
\end{document}


\preprint{APS/123-QED}

\title{Supplementary Information for ``Dynamic Mode Decomposition for Extrapolating Non-equilibrium Green's Functions Dynamics"}
\maketitle
\noindent\hspace{-.02\textwidth}\begin{minipage}{\textwidth}
 \centering
\begin{minipage}{\textwidth}
    \includegraphics[width = \textwidth]{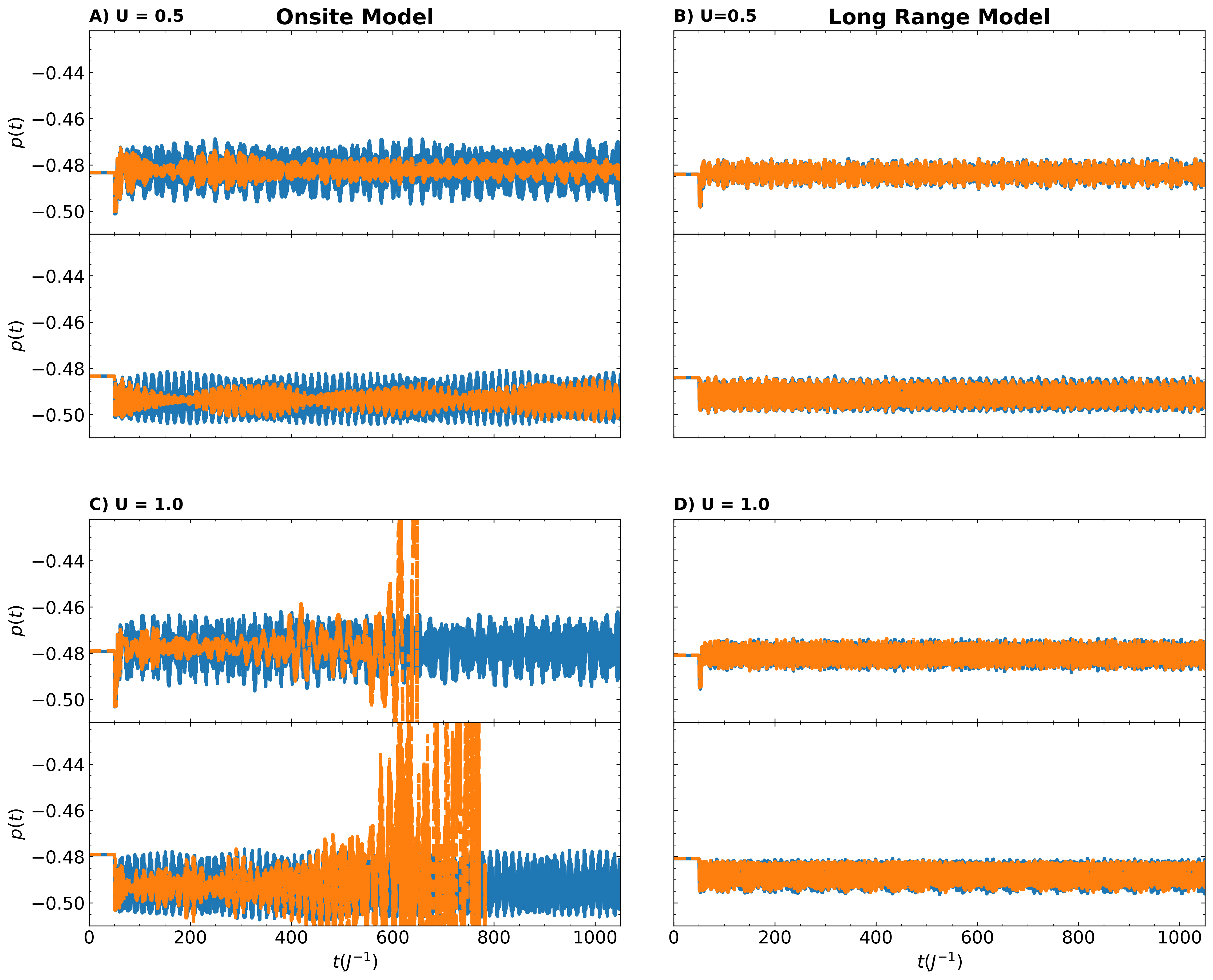}
    \captionof{figure}{Full trajectory for 8 site model. Corresponding to Fig. 2 in the main text.}
    \label{fig:full_trajectory}
\end{minipage}%
\end{minipage}
\begin{figure*}[h!]
    \centering
    \includegraphics[width=\textwidth]{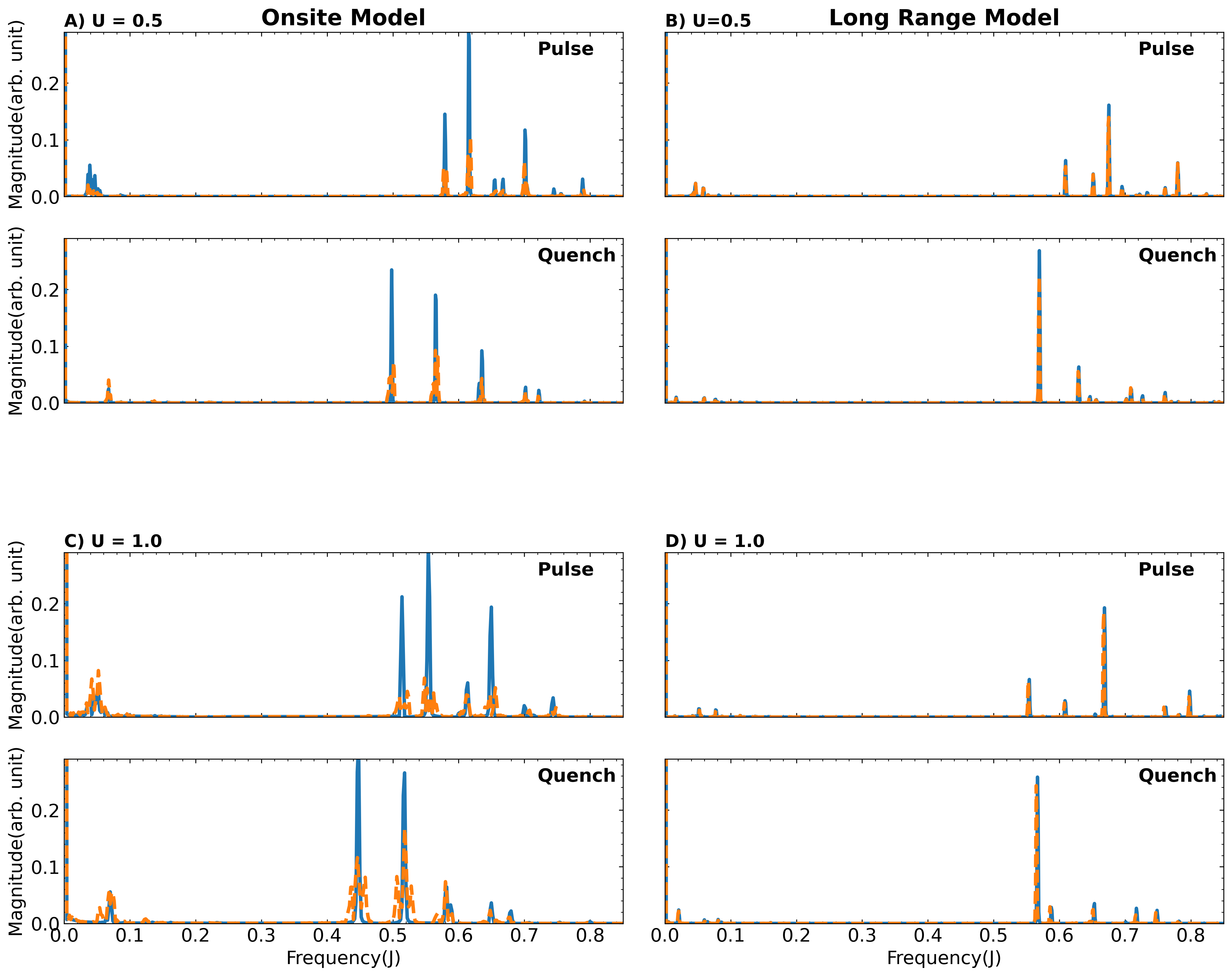}
    \caption{Full frequency spectrum for 8 site model.  Corresponding to Fig. 3 in the main text}
    \label{SI:full_spectrum}
\end{figure*}
\begin{figure*}
    \centering
    \includegraphics[width = \textwidth]{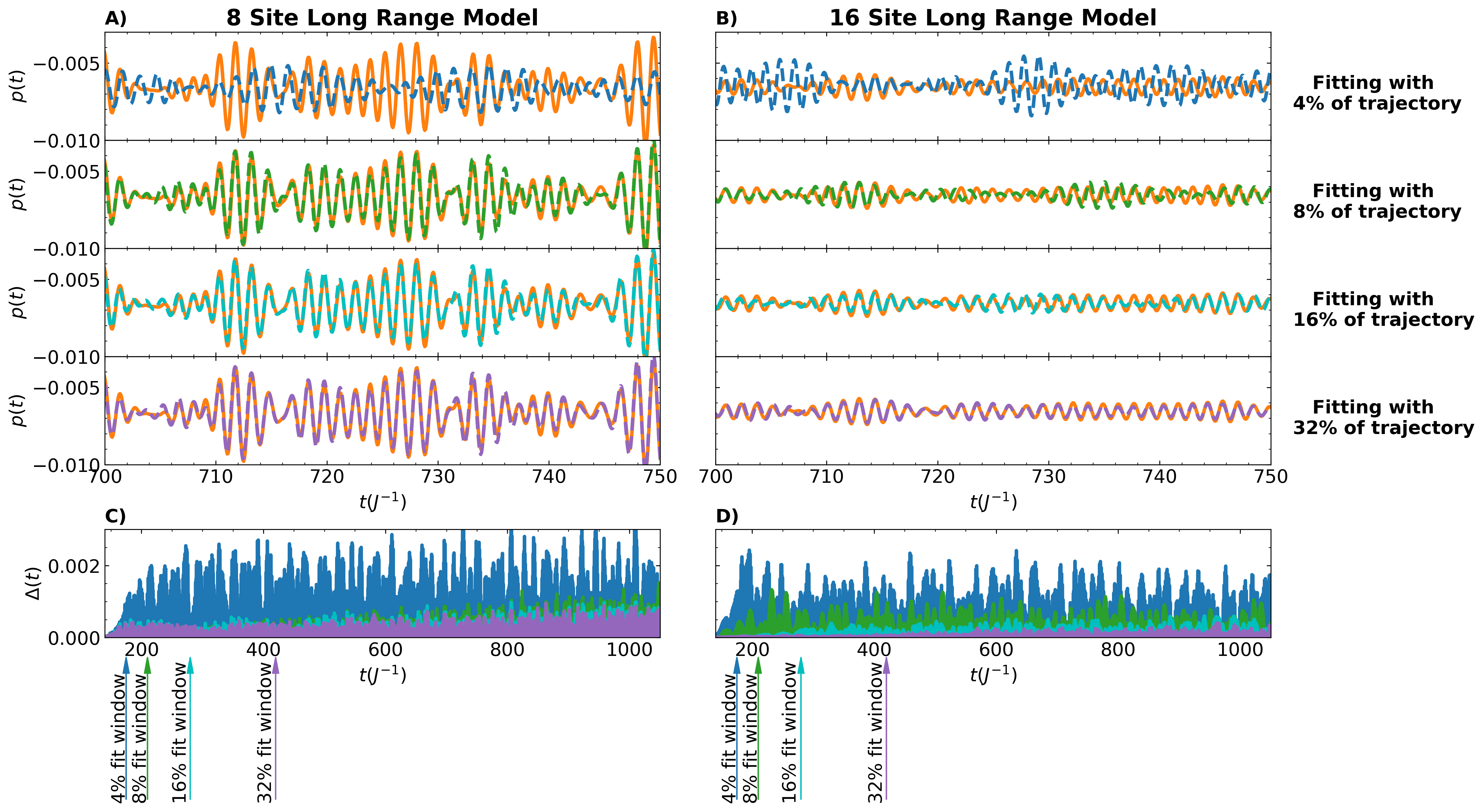}
    \caption{DMD result for $G_{15}$ for 8 and 16 site model with pulse quench.  }
    \label{fig:G_15}
\end{figure*}
\begin{figure*}
    \centering
    \includegraphics[width = \textwidth]{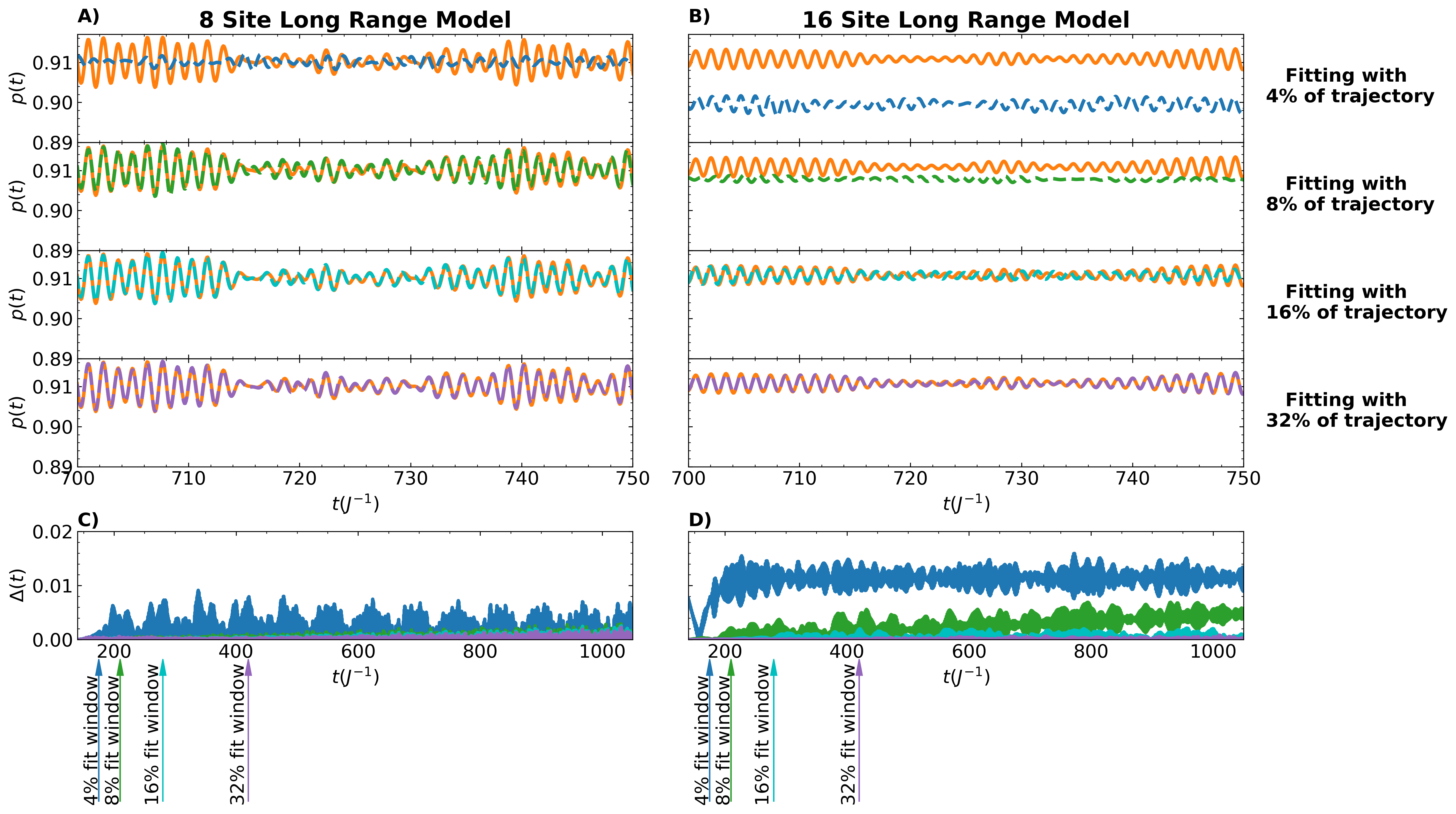}
    \caption{DMD result for $G_{33}$ for 8 and 16 site model with pulse quench}
    \label{fig:G_33}
\end{figure*}
\begin{figure*}
    \centering
    \includegraphics[width = \textwidth]{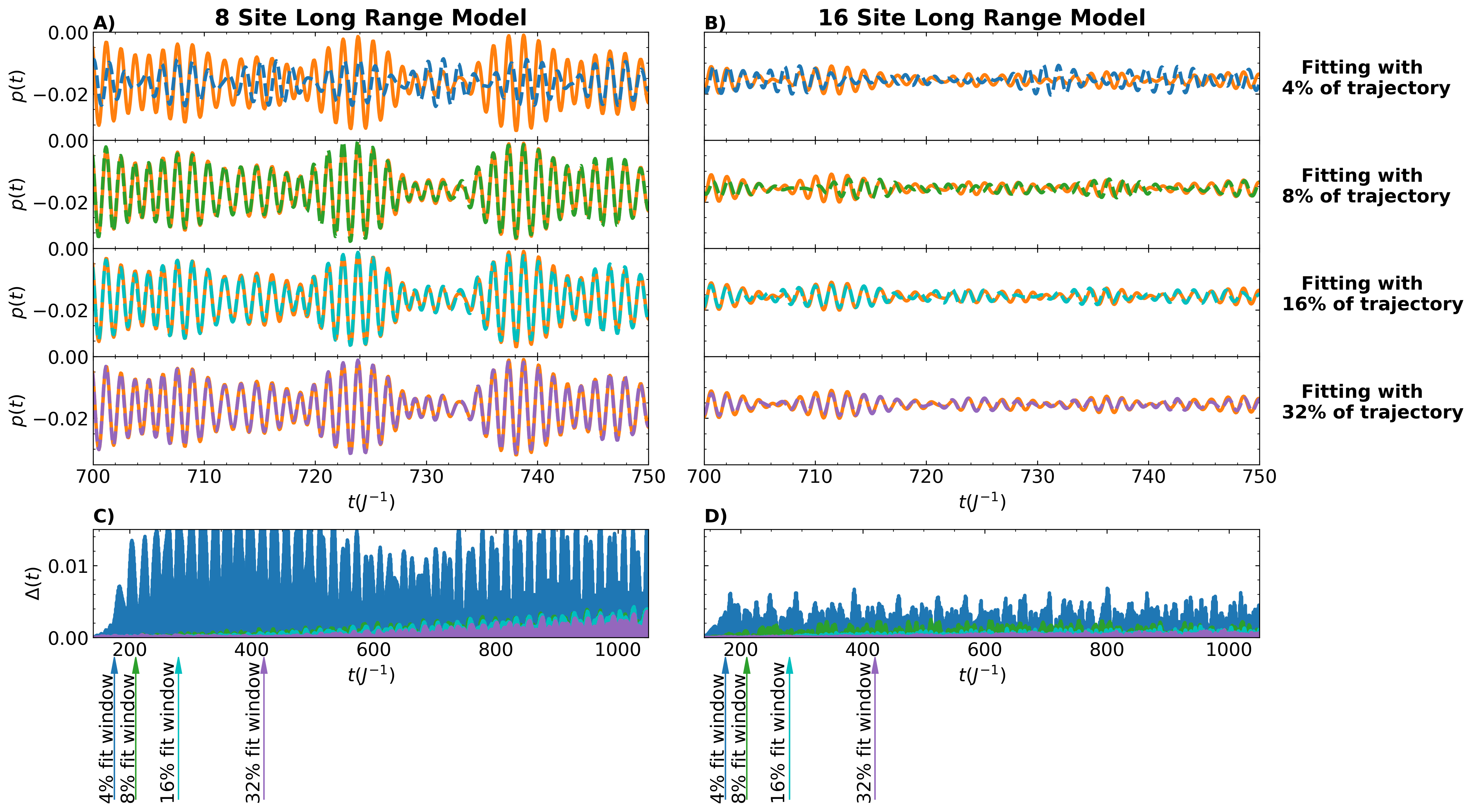}
    \caption{DMD result for $G_{36}$ for 8 and 16 site model with pulse quench}
    \label{fig:G_36}
\end{figure*}
\begin{figure*}
    \centering
    \includegraphics[width = \textwidth]{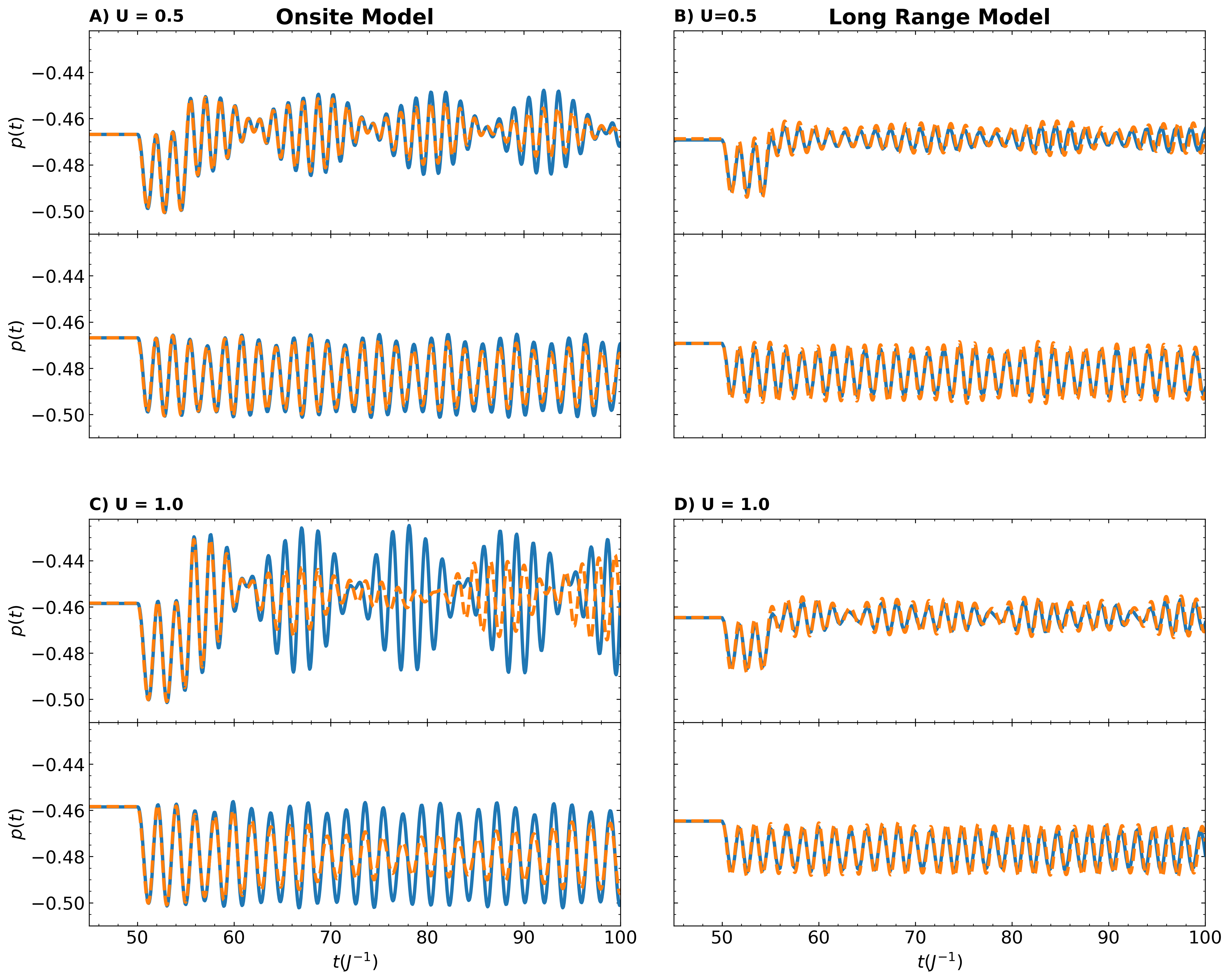}
    \caption{Early time dynamics of 4 site model with onsite interactions and long range interactions with $\frac{1}{r}$ decay.}
    \label{fig:1_r}
\end{figure*}